\documentclass[aps,preprint]{revtex4-1}
\usepackage{graphicx}
\usepackage{amsmath}
\usepackage{makeidx}
\usepackage{amsfonts}
\usepackage{amssymb}
\usepackage{floatrow}

\begin{document}

\title{Statistical physics of loopy interactions: Independent-loop approximation and beyond}

\author{A. Ramezanpour}
\email{aramezanpour@gmail.com}
\affiliation{Department of Physics, University of Neyshabur, P.O. Box 91136-899, Neyshabur, Iran}

\author{S. Moghimi-Araghi}
\email{samanimi@sharif.edu}
\affiliation{Department of Physics, Sharif University of Technology, P.O. Box 11155-9161, Tehran, Iran}

\date{\today}

\begin{abstract}
We consider an interacting system of spin variables on a loopy interaction graph, identified by a tree graph and a set of loopy interactions. We start from a high-temperature expansion for loopy interactions represented by a sum of nonnegative contributions from all the possible frustration-free loop configurations. We then compute the loop corrections using different approximations for the nonlocal loop interactions induced by the spin correlations in the tree graph. For distant loopy interactions, we can exploit the exponential decay of correlations in the tree interaction graph to compute loop corrections within an independent-loop approximation. Higher orders of the approximation are obtained by considering the correlations between the nearby loopy interactions involving larger number of spin variables. In particular, the sum over the loop configurations can be computed "exactly" by the belief propagation algorithm in the low orders of the approximation as long as the loopy interactions have a tree structure. These results might be useful in developing more accurate and convergent message-passing algorithms exploiting the structure of loopy interactions.
\end{abstract}

%\pacs{05.50.+q,75.50.Lk} 

\maketitle

\section{Introduction}\label{S0}
The problem of computing local marginals of an arbitrary probability measure is computationally hard but essential, for example, in the study of inverse problems and in solving for solutions to random constraint satisfaction problems. The loopy belief propagation (BP) algorithm is an efficient approximate algorithm that has proven very helpful in the study of random optimization problems \cite{MP-epjb-2001,KFL-inform-2001,MZ-pre-2002,MPZ-science-2002,MM-book-2009}. The BP algorithm, relying on the Bethe approximation, is exact for systems living on tree interaction graphs. It is also expected to be asymptotically exact for locally tree-like graphs as long as the variables are not strongly correlated. 
In general, the accuracy and convergence of the loopy BP algorithm are not guaranteed, especially in the presence of short loopy interactions. Therefore, characterizing the algorithm performance in the presence of loopy interactions  \cite{Y-2002,H-2006,WYM-2012}, and its generalizations \cite{P-jphysa-2005,YFW-nips-2001,MQ-2004,WJW-ieee-2003,K-ieee-2006}, have been the subject of many studies in recent years. In fact, the loopy BP marginals are not globally consistent in the presence of loopy interactions. This means that the algorithm performance can be improved by demanding more consistency for the BP marginals, e.g., by ensuring that the local marginals satisfy the fluctuation-response relations \cite{MR-jstat-2006,MM-jpp-2009,YT-pre-2013}.
There are also many efforts to take into account more accurately the effect of loopy interactions by an expansion around the BP solution in a loopy graph \cite{E-physica-1990,PS-jstat-2006,CC-pre-2006,CC-jstat-2006,MWKR-proc-2007,SWW-pnip-2007,F-jstat-2013,R-pre-2013}.      

\begin{figure}
\includegraphics[width=4cm]{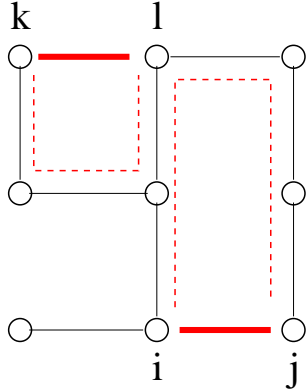}
\caption{(Color online) A loopy interaction graph consisting of a spanning tree and two loopy interactions $(ij)$ and $(kl)$ of lengths $l_{ij}=5$ and $l_{kl}=3$, respectively. The two loops have distance $d_{ij,kl}=2$ defined as the minimum separation of their endpoints along the tree.}\label{f1}
\end{figure}

In this study, we consider the Ising model on a loopy interaction graph, e.g., the two-dimensional square lattice. Given a spanning tree of the interaction graph, we can unambiguously identify the set of loopy interactions, the number of links between the endpoints of a loopy interaction on the spanning tree (length of loop), and the minimum number of links between the endpoints of two loopy interactions (distance of two loops) (see Fig. \ref{f1}). We write a high-temperature expansion for the loopy interactions that is given by a sum over all the possible frustration-free loop configurations. A subset of loopy interactions is called frustration-free if there exists at least one spin configuration that satisfies all the interactions in that subset. We consider the loop expansion as a partition function for a system of globally interacting loop variables. The nonlocal loop interactions can be expressed in terms of the correlation functions of the spin variables in the tree interaction graph. We then resort to approximations of different complexity to study such a system of globally interacting loop variables. See also Ref. \cite{F-jstat-2013}, where a high-temperature expansion of loopy interactions was proposed, and the low orders of loop corrections to the Gibbs free energy were computed.         

The following arguments rely heavily on the clustering property of the pure Gibbs state defined by the probability measure of the spin variables on the tree interaction graph: the fact that spin correlations decay exponentially with distance along the tree. Therefore, as long as the loops are well separated, we can approximate the system of loopy interactions by an effective system of independent loops. This independent-loop (IL) approximation works reasonably well in the high-temperature phase down to the critical region of the ferromagnetic Ising model in two spatial dimensions. Here, the approximation performance is determined by the length and distance distribution of the loopy interactions; in one hand, short loops have more contribution in the loop expansion than the longer ones, and in the other hand, the more distant loops are better described within the IL approximation.               

To go beyond the IL approximation, we make use of the Bethe approximation to write the global loop interactions in terms of local correlation functions involving the neighboring loopy interactions. These local correlations in the tree graph can be computed exactly, and then the loop corrections can be computed by the standard belief propagation (BP) algorithm. In this way, and in the low orders of the approximation, we obtain results that are comparable to those of the loopy BP, with algorithms that exhibit better convergence properties than the loopy BP algorithm. This is because these approximations rely on the structure of the loopy interactions instead of the whole interaction graph. Indeed for the Ising model with random couplings on the two-dimensional square lattice, already the IL approximation works better than the loopy BP in estimating the spin correlations on loopy interactions.  

The paper is organized as follows. Section \ref{S1} gives the basic definitions and equations we use in this paper.
In Sec. \ref{S2}, we write the main equations for the high-temperature loop expansions. In Sec. \ref{S3}, we present approximate algorithms based on the loop expansions, starting from the independent-loop approximation.
Finally, Sec. \ref{S4} gives the concluding remarks.

\section{Basic definitions and equations}\label{S1}
Consider the Ising model on an arbitrary graph $G$ with Hamiltonian: 
\begin{align}
H[\boldsymbol\sigma]=-\sum_{i=1}^N h_i \sigma_i-\sum_{(ij) \in G} J_{ij} \sigma_i\sigma_j,
\end{align}
where $\boldsymbol\sigma=\{\sigma_i|i=1,\dots,N\}$ with $\sigma_i \in \{-1,+1\}$.
The probability of finding the system in spin configuration $\boldsymbol\sigma$ at inverse temperature $\beta=1/T$ is given by the Gibbs probability measure,
\begin{align}
\mu(\boldsymbol\sigma)=\frac{1}{Z} e^{-\beta H[\boldsymbol\sigma]}.
\end{align}
The partition function $Z$ and the free energy $F$ are given by
\begin{align}
Z=\sum_{\boldsymbol\sigma} e^{-\beta H[\boldsymbol\sigma]}=e^{-\beta F}.
\end{align}

\begin{figure}
\includegraphics[width=16cm]{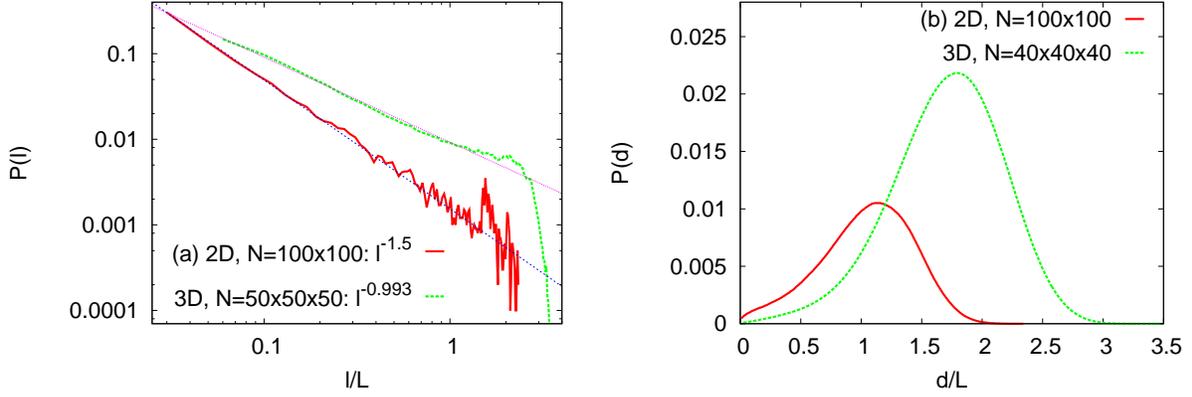}
\caption{(Color online) (a) Length and (b) distance probability distributions $P(l), P(d)$ of the loopy interactions in the two-dimensional ($2D$) and three-dimensional ($3D$) square lattices with respect to a random spanning tree. Here $L$ is the linear size of the lattices with periodic boundary conditions.}\label{f2}
\end{figure}

We choose a spanning tree $\mathsf{T}$ of the interaction graph $G$. This defines the set of loopy interactions $\mathcal{L}=\{(ij) \in G\setminus \mathsf{T}\}$. 
The length $l_{ij}$ of a loopy interaction $(ij)\in \mathcal{L}$ is defined by the number of links connecting the endpoints $i$ and $j$ on the spanning tree. The distance $d_{ij,kl}$ of two loopy interactions is defined by the minimum number of links connecting the endpoints $(i,j)$ to $(k,l)$ on the spanning tree, i.e., $d_{ij,kl}=\min\{l_{ik},l_{il},l_{jk},l_{jl}\}$. Figure \ref{f2} shows the length and distance distributions of loops in two- and three-dimensional cubic lattices.   
We use $\partial i, \partial_0 i$ and $\bar{\partial}_0 i$ for the neighborhood set of $i$ in $G, \mathsf{T}$ and $\mathcal{G}\equiv G\setminus \mathsf{T}$, respectively. 

Let us rewrite the Hamiltonian as $H[\boldsymbol\sigma]=H_0[\boldsymbol\sigma]+H_{\epsilon}[\boldsymbol\sigma]$, separating the energy contribution of interactions on the spanning tree and the loopy interactions,
\begin{align}
H_0[\boldsymbol\sigma] &=-\sum_{i=1}^N h_i \sigma_i-\sum_{(ij) \in \mathsf{T}} J_{ij} \sigma_i\sigma_j,\\
H_{\epsilon}[\boldsymbol\sigma] &=-\sum_{(ij) \in \mathcal{L}} J_{ij} \sigma_i\sigma_j,
\end{align}
and define
\begin{align}
Z_0 =\sum_{\boldsymbol\sigma} e^{-\beta H_0[\boldsymbol\sigma]}=e^{-\beta F_0}.
\end{align}

For the model on the spanning tree $\mathsf{T}$, we can exactly compute the local marginals 
by the Bethe equations for the cavity marginals $\mu_{i\to j}^0(\sigma_i)$. Here, $\mu_{i\to j}^0(\sigma_i)$ is the probability of finding spin $i$ in state $\sigma_i$, in the absence of interaction with spin $j$, that is,
\begin{align}
\mu_{i\to j}^0(\sigma_i) \propto e^{B_i \sigma_i}\prod_{k\in \partial_0 i\setminus j}\left(\sum_{\sigma_k} e^{K_{ik}\sigma_i\sigma_k} \mu_{k\to i}^0(\sigma_k)\right).
\end{align}
These are the belief propagation (BP) equations \cite{KFL-inform-2001,MM-book-2009}. To shorten the notation, we defined $B_i=\beta h_i$ and $K_{ij}=\beta J_{ij}$.

There is only one solution to the above equations which can be obtained by iteration starting from arbitrary cavity marginals.    
Then, the local one- and two-spin marginals are given by
\begin{align}
\mu_{i}^0(\sigma_i) &\propto e^{B_i \sigma_i}\prod_{k\in \partial_0 i}\left(\sum_{\sigma_k} e^{K_{ik}\sigma_i\sigma_k} \mu_{k\to i}^0(\sigma_k)\right),\\
\mu_{ij}^0(\sigma_i,\sigma_j) &\propto  e^{K_{ij}\sigma_i\sigma_j} \mu_{i\to j}^0(\sigma_i)\mu_{j\to i}^0(\sigma_j).
\end{align}
At the end, the free energy is obtained by $F_0=\sum_i \Delta F_i^0-\sum_{(ij)\in \mathsf{T}} \Delta F_{ij}^0$, where the local free energy changes read as
\begin{align}
e^{-\beta \Delta F_i^0} &=\sum_{\sigma_i} e^{B_i\sigma_i}\prod_{j\in \partial_0 i}\left(\sum_{\sigma_j} e^{K_{ij}\sigma_i\sigma_j} \mu_{j\to i}^0(\sigma_j)\right),\\
e^{-\beta \Delta F_{ij}^0} &=\sum_{\sigma_i,\sigma_j}e^{K_{ij}\sigma_i\sigma_j}\mu_{i\to j}^0(\sigma_i)\mu_{j\to i}^0(\sigma_j).
\end{align}
The reader can refer to Ref. \cite{MM-book-2009} for derivation of the above equations. 

The above marginals are enough to obtain the local expectation values $m_i^0=\left\langle \sigma_i\right\rangle_0, c_{ij}^0=\left\langle \sigma_i\sigma_j\right\rangle_0$. We can use the Bethe formalism to compute the higher order correlations $\left\langle \prod_{i}\sigma_i^{x_i}  \right\rangle_0$ for $x_i\in \{0,1\}$, 
\begin{align}
\left\langle \prod_{i}\sigma_i^{x_i}  \right\rangle_0 =\frac{1}{Z_0}\sum_{\boldsymbol\sigma} \prod_{i}\sigma_i^{x_i} e^{-\beta H_0[\boldsymbol\sigma]}=e^{\sum_i \Delta_i-\sum_{(ij)\in \mathsf{T}}\Delta_{ij}},
\end{align}
where
\begin{align}
e^{\Delta_i} &=\frac{\sum_{\sigma_i} \sigma_i^{x_i}e^{B_i\sigma_i}\prod_{j\in \partial_0 i}\left(\sum_{\sigma_j} e^{K_{ij}\sigma_i\sigma_j} \nu_{j\to i}^0(\sigma_j)\right)}{\sum_{\sigma_i} e^{B_i\sigma_i} \prod_{j\in \partial_0 i}\left(\sum_{\sigma_j} e^{K_{ij}\sigma_i\sigma_j} \mu_{j\to i}^0(\sigma_j)\right)},\\
e^{\Delta_{ij}} &=\frac{\sum_{\sigma_i,\sigma_j}e^{K_{ij}\sigma_i\sigma_j}\nu_{i\to j}^0(\sigma_i)\nu_{j\to i}^0(\sigma_j)}{\sum_{\sigma_i,\sigma_j}e^{K_{ij}\sigma_i\sigma_j}\mu_{i\to j}^0(\sigma_i)\mu_{j\to i}^0(\sigma_j)}.
\end{align}
The $\mu_{i\to j}^0(\sigma_i)$ are the BP cavity marginals given before, and the new cavity marginals satisfy the following BP equations:
\begin{align}
\nu_{i\to j}^0(\sigma_i) \propto \sigma_i^{x_i}e^{B_i \sigma_i}\prod_{k\in \partial_0 i\setminus j}\left(\sum_{\sigma_k} e^{K_{ik}\sigma_i\sigma_k} \nu_{k\to i}^0(\sigma_k)\right).
\end{align}
Along the same lines, one can compute the probability of observing a subset of spin variables  in a given configuration, $\left\langle \prod_{i\in X}\delta_{\sigma_i,\sigma_i^*} \right\rangle_0$.

The extension of the above equations to loopy graphs gives the approximate loopy BP equations; this is simply done by replacing $\partial_0 i$ with $\partial i$ in the above equations.

\section{A high-temperature expansion of the loopy interactions}\label{S2}
Let us expand the partition function $Z$ for small couplings in $H_{\epsilon}$ to get
\begin{multline}\label{L0}
e^{-\beta(F-F_0)}=\left\langle \prod_{(ij) \in \mathcal{L}}\left(\cosh K_{ij}+\sinh K_{ij} \sigma_i\sigma_j \right) \right\rangle_0 \\ =\sum_{\mathbf{s}} \prod_{(ij) \in \mathcal{L}} u_{ij}(s_{ij}) \left\langle \prod_{(ij) \in \mathcal{L}}(\sigma_i\sigma_j)^{s_{ij}}  \right\rangle_0,
\end{multline}
where $\mathbf{s}=\{s_{ij}|(ij) \in \mathcal{L}\}$ identifies a loop configuration with $s_{ij}\in \{0,1\}$ showing the absence or presence of an interaction.
Here $u_{ij}(s_{ij})\equiv \delta_{s_{ij},0}\cosh K_{ij}+\delta_{s_{ij},1}\sinh K_{ij}$. The averages
\begin{align}
\left\langle \prod_{(ij) \in \mathcal{L}}(\sigma_i\sigma_j)^{s_{ij}}  \right\rangle_0=\frac{1}{Z_0}\sum_{\boldsymbol\sigma} e^{-\beta H_0[\boldsymbol\sigma]} \prod_{(ij) \in \mathcal{L}}(\sigma_i\sigma_j)^{s_{ij}},
\end{align}
can be computed efficiently, as described in the previous section. Note that these averages are trivial when all the nodes in the subgraph $\mathcal{G}(\mathbf{s})$, induced by the present loopy interactions in $\mathbf{s}$, have an even degree.

We rewrite the right hand side of the above expansion in another form as 
\begin{align}\label{L1}
e^{-\beta(F-F_0)}=e^{\sum_{(ij) \in \mathcal{L}}K_{ij}}\sum_{\mathbf{s}} \prod_{(ij) \in \mathcal{L}} (2e^{-K_{ij}}\sinh K_{ij})^{s_{ij}} \left\langle \prod_{(ij) \in \mathcal{L}}(\frac{\sigma_i\sigma_j-1}{2})^{s_{ij}}  \right\rangle_0,
\end{align}
with $(\frac{\sigma_i\sigma_j-1}{2})^0=1$ independent of the value of $\sigma_i\sigma_j$.  
The leading term $e^{\sum_{(ij) \in \mathcal{L}}K_{ij}}$ is obtained when all the $s_{ij}$ are zero. In this way, we obtain a high-temperature expansion for the loopy interactions:
\begin{align}\label{L2}
e^{-\beta(F-F_0-\Delta F_0)}= \sum_{\mathbf{s}} \prod_{(ij) \in \mathcal{L}} v_{ij}^+(s_{ij}) \left\langle \prod_{(ij) \in \mathcal{L}}(\frac{1-\sigma_i\sigma_j}{2})^{s_{ij}}  \right\rangle_0 \equiv e^{-\beta \Delta F^+},
\end{align}
where now
\begin{align}\label{Leven}
-\beta \Delta F_0 \equiv \sum_{(ij) \in \mathcal{L}}K_{ij},\hskip1cm
v_{ij}^+(s_{ij}) \equiv  (e^{ -2K_{ij}}-1)^{s_{ij}}.
\end{align}

Let us define $M_+(\mathbf{s})\equiv \sum_{(ij) \in \mathcal{L}}s_{ij}\theta(K_{ij})$ as the number of ferromagnetic loopy interactions in loop configuration $\mathbf{s}$. The step function $\theta(K_{ij})=1$ for ferromagnetic loopy interactions ($K_{ij}>0$), otherwise $\theta(K_{ij})=0$. 
This allows us to separate the positive and negative contributions to the expansion,
\begin{multline}\label{L3}
e^{-\beta(F-F_0-\Delta F_0)}= \sum_{\mathbf{s}:M_+=even} \prod_{(ij) \in \mathcal{L}} |v_{ij}^+(s_{ij})| \left\langle \prod_{(ij) \in \mathcal{L}}(\frac{1-\sigma_i\sigma_j}{2})^{s_{ij}}  \right\rangle_0 \\
-\sum_{\mathbf{s}:M_+=odd} \prod_{(ij) \in \mathcal{L}} |v_{ij}^+(s_{ij})| \left\langle \prod_{(ij) \in \mathcal{L}}(\frac{1-\sigma_i\sigma_j}{2})^{s_{ij}}  \right\rangle_0.
\end{multline}
We know that the right hand side is positive, and if we are interested in the free energy densities $f\equiv F/N$ in the thermodynamic limit, then 
\begin{align}
f &=f_0+\Delta f_0+\Delta f_{even}^+,\\
e^{-\beta \Delta F_{even}^+} &= \sum_{\mathbf{s}:M_+=even} \prod_{(ij) \in \mathcal{L}} |v_{ij}^+(s_{ij})| \left\langle \prod_{(ij) \in \mathcal{L}}(\frac{1-\sigma_i\sigma_j}{2})^{s_{ij}}  \right\rangle_0.
\end{align}
In other words, it is $\Delta f_{even}^+$ that is responsible for any nonanalytic behavior of the free energy density.

Note that instead of working with the ferromagnetic loopy interactions, we could work with the anti-ferromagnetic ones to get
\begin{align}\label{L4}
e^{-\beta(F-F_0+\Delta F_0)}=\sum_{\mathbf{s}} \prod_{(ij) \in \mathcal{L}}  v_{ij}^-(s_{ij}) \left\langle \prod_{(ij) \in \mathcal{L}}(\frac{1+\sigma_i\sigma_j}{2})^{s_{ij}}  \right\rangle_0 \equiv e^{-\beta \Delta F^-},
\end{align}
where $v_{ij}^-(s_{ij})\equiv (e^{2K_{ij}}-1)^{s_{ij}}$. Here the positive contribution to the expansion comes from loop configurations with an even number of anti-ferromagnetic loopy interactions $M_-(\mathbf{s})\equiv \sum_{(ij) \in \mathcal{L}}s_{ij}\theta(-K_{ij})$. Thus, in the thermodynamic limit, we have
\begin{align}
f &=f_0-\Delta f_0+\Delta f_{even}^-,\\
e^{-\beta \Delta F_{even}^-} &\equiv \sum_{\mathbf{s}:M_-=even} \prod_{(ij) \in \mathcal{L}} |v_{ij}^-(s_{ij})| \left\langle \prod_{(ij) \in \mathcal{L}}(\frac{1+\sigma_i\sigma_j}{2})^{s_{ij}}  \right\rangle_0.
\end{align}

The two different loop expansions provide the same free energy, therefore, 
\begin{align}
\Delta F_0+\Delta F^+ &=-\Delta F_0+\Delta F^-,\\
\Delta f_0+\Delta f_{even}^+ &=-\Delta f_0+\Delta f_{even}^-.
\end{align}
These are indeed relationships between the free energies of two interacting systems of loop variables, with effective energy functions $\mathcal{H}^{\pm}[\mathbf s]$ that depend on the couplings of the original system defined by $H[\boldsymbol \sigma]$. And, in the presence of a phase transition, the two systems display the same critical behavior. 

Finally, we can use both the ferromagnetic and anti-ferromagnetic representations to write the loop expansion as
\begin{align}\label{L5}
e^{-\beta(F-F_0-\Delta F_0)}=\sum_{\mathbf{s}} \prod_{(ij) \in \mathcal{L}}  v_{ij}(s_{ij}) \left\langle \prod_{(ij) \in \mathcal{L}}\left(\frac{1-\xi_{ij}\sigma_i\sigma_j}{2}\right)^{s_{ij}}  \right\rangle_0.
\end{align}
Now $\beta \Delta F_0 \equiv \sum_{(ij) \in \mathcal{L}} |K_{ij}|$, $v_{ij}(s_{ij}) \equiv  (e^{ 2|K_{ij}|}-1)^{s_{ij}}$, and we defined $\xi_{ij}\equiv (-1)^{\theta(K_{ij})}$. Note that here every loop configuration $\mathbf{s}$ has a nonnegative contribution to the expansion. The average $\left\langle \prod_{(ij) \in \mathcal{L}}(\frac{1-\xi_{ij} \sigma_i\sigma_j}{2})^{s_{ij}}  \right\rangle_0$ is the probability of having $x_{ij}\equiv (\frac{1-\xi_{ij} \sigma_i\sigma_j}{2})^{s_{ij}}=1$ for all the loopy interactions. We denote this probability by $P_0\left(\prod _{(ij)\in \mathcal{L}} x_{ij}=1\right)$. The complexity of computing this probability grows exponentially with the number of connected clusters in the graph $\mathcal{G}(\mathbf{s})$ induced by the set of present loopy interactions in $\mathbf{s}$; the neighboring spins in each cluster should satisfy $K_{ij}\sigma_i\sigma_j>0$, and this is possible only if each cluster of loopy interactions is free of frustration, otherwise $P_0\left(\prod _{(ij)\in \mathcal{L}} x_{ij}=1\right)=0$. In other words, only frustration-free loop configurations enter in the above expansion, and each configuration has a nonnegative contribution.

\section{Approximating the nonlocal loop interactions}\label{S3}
So far, we have made no approximation in deriving the high-temperature loop expansions presented in the previous section. But, to obtain the loop corrections, still we have to compute the nonlocal loop interactions that appear in these expansions. In this section, we will focus on different approximations to simplify the loop interactions.

\subsection{Independent-loop approximation}\label{S31}
For tree interaction graphs, we are sure that the averages $\left\langle \prod_{(ij) \in \mathcal{L}}(\sigma_i\sigma_j)^{s_{ij}}  \right\rangle_0$ are computed within a pure state, where clustering holds and with a good approximation for distant loopy interactions      
\begin{align}
\left\langle \prod_{(ij) \in \mathcal{L}}(\sigma_i\sigma_j)^{s_{ij}}  \right\rangle_0 \approx \prod_{(ij) \in \mathcal{L}} \left\langle (\sigma_i\sigma_j)^{s_{ij}}  \right\rangle_0.
\end{align}
This independent-loop (IL) approximation results in
\begin{align}
e^{-\beta(F-F_0)} \approx \prod_{(ij) \in \mathcal{L}} [\cosh K_{ij}+\sinh K_{ij}\left\langle \sigma_i\sigma_j \right\rangle_0]. 
\end{align}
Note that in a ferromagnetic system this approximation underestimates the correlations and thus provides an upper bound for the free energy.

In a homogeneous system $K_{ij}=K$, and in the absence of external fields $B_i=0$, we have $\langle \sigma_i\sigma_j \rangle_0=\tanh^{l_{ij}}(K)$, thus
\begin{align}
-\beta F\approx -\beta F_0 + \sum_l n_l \ln\left(\cosh(K)+\sinh(K)\tanh^{l}(K)\right), 
\end{align}
where $n_l$ is the number of loopy interactions of length $l$.
  
In general, we obtain the magnetization by
\begin{align}
m_i=\left\langle \sigma_i \right\rangle=-\frac{\partial \beta F}{\partial B_i} \approx \left\langle \sigma_i \right\rangle_0 + \sum_{(kl) \in \mathcal{L}} \frac{\sinh K_{kl}[\left\langle \sigma_k\sigma_l\sigma_i \right\rangle_0-\left\langle \sigma_k\sigma_l\right\rangle_0 \left\langle \sigma_i \right\rangle_0]}{\cosh K_{kl}+\sinh K_{kl}\left\langle \sigma_k\sigma_l \right\rangle_0}, 
\end{align}
and the two-spin correlation $c_{ij}=\left\langle \sigma_i\sigma_j \right\rangle=-\frac{\partial \beta F}{\partial K_{ij}}$ by
\begin{align}
c_{ij} &\approx \frac{\sinh K_{ij}+\cosh K_{ij}\left\langle \sigma_i\sigma_j \right\rangle_0}{\cosh K_{ij}+\sinh K_{ij}\left\langle \sigma_i\sigma_j \right\rangle_0}, \hskip0.5cm (ij)\in \mathcal{L}\\
c_{ij} &\approx \left\langle \sigma_i\sigma_j \right\rangle_0 +\sum_{(kl) \in \mathcal{L}} \frac{\sinh K_{kl}[\left\langle \sigma_k\sigma_l\sigma_i\sigma_j \right\rangle_0-\left\langle \sigma_k\sigma_l\right\rangle_0 \left\langle \sigma_i\sigma_j \right\rangle_0]}{\cosh K_{kl}+\sinh K_{kl}\left\langle \sigma_k\sigma_l \right\rangle_0}, \hskip0.5cm (ij)\notin \mathcal{L}. 
\end{align}
 
\begin{figure}
\includegraphics[width=16cm]{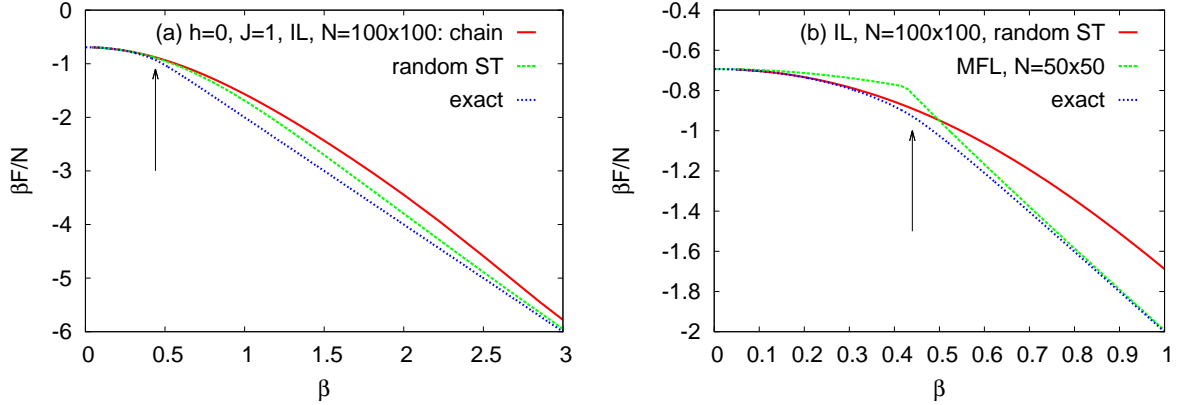}
\caption{(Color online) (a) The exact free energy of the ferromagnetic Ising model ($h_i=0,J_{ij}=1$) on the $2D$ square lattice is compared with the independent-loop approximation based on a random and regular (chain) spanning tree. (b) The IL approximation compared with a mean-field approximation of loopy interactions (MFL), where a loopy interaction $(ij)\in \mathcal{L}$ is replaced with effective fields $h_i^{(ij)}=\frac{1}{2}J_{ij}m_j$ and $h_j^{(ij)}=\frac{1}{2}J_{ij}m_i$ determined self-consistently. The arrows show the position of the ferromagnetic phase transition.}\label{f3}
\end{figure}

In Fig. \ref{f3} we show the free energy obtained by the IL approximation for the ferromagnetic Ising model on the two-dimensional square lattice.    
We also compare the IL approximation with a mean-field approximation of the loopy interactions \cite{R-pre-2013}, where a loopy interaction $J_{ij}\sigma_i\sigma_j$ is replaced with two effective fields $h_i^{(ij)}=\frac{1}{2}J_{ij}m_j$ and $h_j^{(ij)}=\frac{1}{2}J_{ij}m_i$ acting on spins $i$ and $j$. Here the local magnetizations $m_{i}$ are determined self-consistently. One can indeed do better than this naive approximation by replacing the $(m_i,m_j)$ in the effective fields with the cavity magnetizations $(m_{i\to j},m_{j\to i})$ computed in the absence of the loopy interaction.     
  
\subsection{Independent-site approximation}\label{S32}
In the independent-loop (IL) approximation we do not consider the overlap of the neighboring loopy interactions. An approximate way to do this is the following:  
\begin{align}
\left\langle \prod_{(ij) \in \mathcal{L}}(\sigma_i\sigma_j)^{s_{ij}}  \right\rangle_0 \approx \prod_{i \in \mathcal{G}} \left\langle \sigma_i^{\sum_{j\in \bar{\partial}_0 i}s_{ij}} \right\rangle_0,
\end{align}
leading to
\begin{align}
e^{-\beta(F-F_0)} \approx \sum_{\mathbf{s}} \prod_{(ij) \in \mathcal{L}} u_{ij}(s_{ij}) \prod_{i \in \mathcal{G}} u_i(s_{\bar{\partial}_0 i}),
\end{align}
where $u_{ij}(s_{ij})\equiv \delta_{s_{ij},0}\cosh K_{ij}+\delta_{s_{ij},1}\sinh K_{ij}$, and
\begin{align}
u_i(s_{\bar{\partial}_0 i}) \equiv 1+(\left\langle \sigma_i  \right\rangle_0-1)\left(\frac{1-(-1)^{\sum_{j\in \bar{\partial}_0 i}s_{ij}}}{2}\right).
\end{align}
This independent-site (IS) approximation, which again relies on the clustering property, is appropriate for systems in external fields where the local magnetizations $\left\langle \sigma_i  \right\rangle_0$ are nonzero.    

\begin{figure}
\includegraphics[width=10cm]{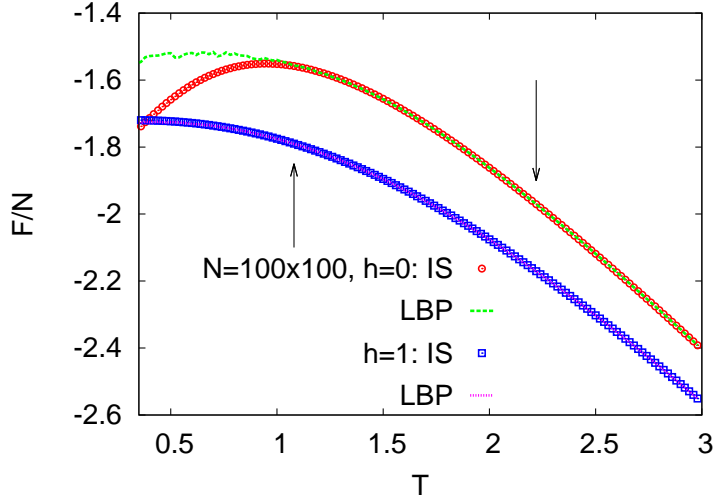}
\caption{(Color online) Comparing the loopy BP (LBP) free energy with that of the independent-site (IS) approximation for a random Ising model ($h_i=\pm h, J_{ij}=\pm 1$) on the $2D$ square lattice. The positive and negative values for the fields and the couplings are chosen with equal probability. The loopy BP algorithm does not converge for $T<T_s$ indicated by the arrows.}\label{f4}
\end{figure}

Now, the sum over the loop configurations can be computed by the Bethe approximation, which is exact when the loopy interactions have a tree structure. Here we need the cavity messages $\lambda_{i\to j}(s_{ij})$ for loop variables $s_{ij}$ in the absence of $u_{ij}$ and $u_j$. These cavity messages are governed by the following Bethe equations:
\begin{align}\label{bp-is}
\lambda_{i\to j}(s_{ij}) \propto \sum_{\{s_{ik}|k\in \bar{\partial}_0 i \setminus j\}} u_i(s_{\bar{\partial}_0 i}) \prod_{k\in \bar{\partial}_0 i \setminus j} \left(u_{ik}(s_{ik})\lambda_{k\to i}(s_{ik})\right).
\end{align}
We solve the above equations by iteration, and compute the free energy from  $F-F_0  \approx \sum_{i\in \mathcal{G}}\Delta F_i-\sum_{(ij)\in \mathcal{L}}\Delta F_{ij}$, where
\begin{align}
e^{-\beta\Delta F_i} &= \sum_{\{s_{ik}|k\in \bar{\partial}_0 i\}} u_i(s_{\bar{\partial}_0 i}) \prod_{k\in \bar{\partial}_0 i} \left(u_{ik}(s_{ik})\lambda_{k\to i}(s_{ik})\right),\\
e^{-\beta\Delta F_{ij}} &= \sum_{s_{ij}}  u_{ij}(s_{ij})\lambda_{j\to i}(s_{ij})\lambda_{i\to j}(s_{ij}).
\end{align}
In Fig. \ref{f4} we compare the above free energy with the exact and loopy BP free energies for random Ising models on the two-dimensional square lattice. We see that in the presence of external fields, the IS free energy is very close to the one obtained by the loopy BP, and at the same time, the IS algorithm converges down to very low temperatures, where the loopy BP does not converge.     

Here the local magnetizations are given by
\begin{align}
m_l=-\frac{\partial \beta F}{\partial B_l} \approx \left\langle \sigma_l \right\rangle_0 - \sum_{i\in \mathcal{G}} \frac{\partial \beta\Delta F_i}{\partial B_l},
\end{align}
which can easily be computed given the $\lambda_{i\to j}(s_{ij})$ and derivatives 
\begin{align}
\frac{\partial u_i(s_{\bar{\partial}_0 i})}{\partial B_l}=\left( \frac{1-(-1)^{\sum_{j\in \bar{\partial}_0 i}s_{ij}}}{2}\right)\left(\left\langle \sigma_i\sigma_l \right\rangle_0-\left\langle \sigma_i \right\rangle_0\left\langle \sigma_l \right\rangle_0 \right).
\end{align}
Note that the Bethe free energy $\sum_{i\in \mathcal{G}}\Delta F_i-\sum_{(ij)\in \mathcal{L}}\Delta F_{ij}$ is stationary with respect to changes in the messages $\lambda_{i\to j}(s_{ij})$, computed at the fixed point of the BP equations. Thus we can safely take 
\begin{align}
\frac{\partial}{\partial \lambda_{i\to j}(s_{ij})}\left(\sum_{i'\in \mathcal{G}}\Delta F_{i'}-\sum_{(i'j')\in \mathcal{L}}\Delta F_{i'j'}\right)\frac{\partial \lambda_{i\to j}(s_{ij})}{\partial B_l}=0.  
\end{align}

\begin{figure}
\includegraphics[width=16cm]{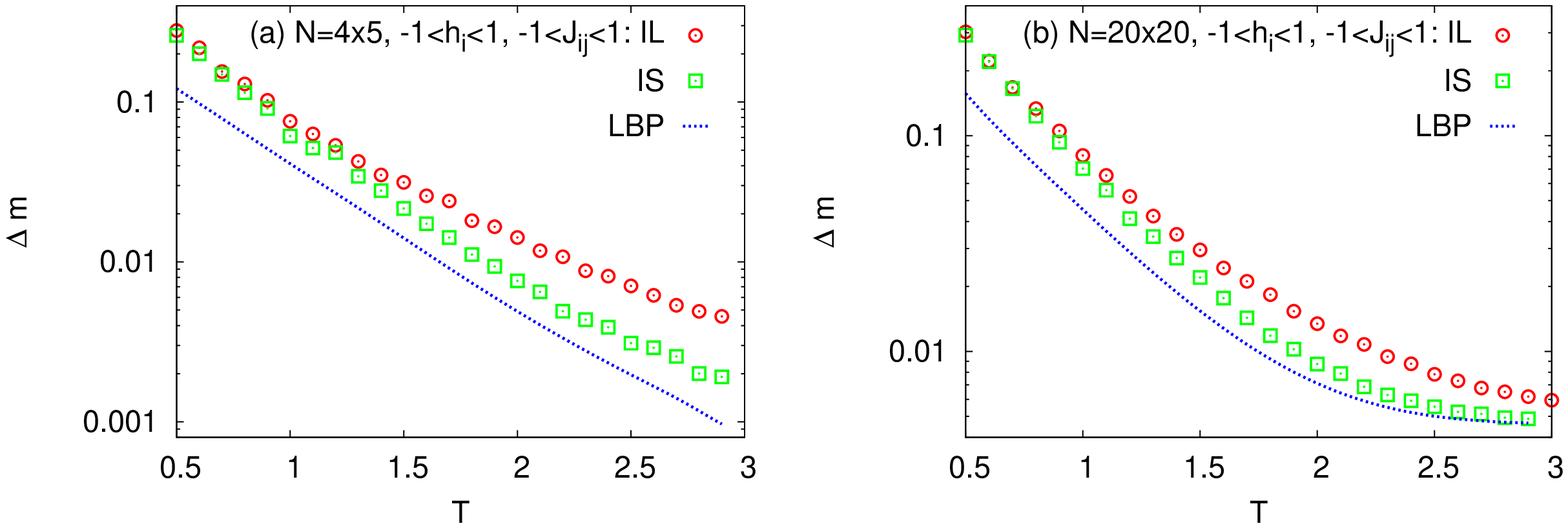}
\caption{(Color online) Comparing the RMS error in magnetization obtained by different approximations: independent-loop (IL), independent-site (IS), and loopy belief propagation (LBP) for a random Ising model with random fields and couplings uniformly distributed in $(-1,+1)$, on the $2D$ square lattice. Both the IL and IS approximations are based on a maximum-weight spanning tree, with weight $W=\sum_{(ij)\in \mathsf{T}} |J_{ij}|$.  
The $m_i^{exact}$ are computed by an exact algorithm (for the smaller system in panel(a)) or by the Monte Carlo algorithm (for the larger system in panel (b)). The errorbars are about the size of the symbols. The RMS error is given by $\Delta m \equiv \sqrt{\sum_{i=1}^N(m_i^{app}-m_i^{exact})^2/N}$.}\label{f5}
\end{figure}

In Appendix \ref{app-mc} we obtain the two-spin correlations $c_{ij}=\left\langle \sigma_i\sigma_j \right\rangle$. Figure \ref{f5} displays the error we make in estimating the local magnetizations by the IL and IS approximations. The figure also shows the results obtained by the loopy BP algorithm, which is still doing better than the above approximations. In the following, we propose a systematic scheme to obtain more accurate approximations.

\subsection{Beyond the independent-loop approximation}\label{S33}
More accurate expressions for the nonlocal loop interactions can be obtained by the Bethe approximation, for example,
\begin{align}
P_0\left(\prod _{(ij)\in \mathcal{L}} x_{ij}=1\right) \approx \prod_{(ij) \in \mathcal{L}} q_{ij}(x_{ij}=1) \prod_{i \in \mathcal{G}}\frac{ q_i(\prod_{j\in \bar{\partial}_0 i} x_{ij}=1) }{\prod_{j\in \bar{\partial}_0 i} q_{ij}(x_{ij}=1)}.
\end{align}
Here, $q_{ij}(x_{ij}=1)$ and $q_i(\prod_{j\in \bar{\partial}_0 i} x_{ij}=1)$ are, respectively, the probability of having $x_{ij}=1$ and $\prod_{j\in \bar{\partial}_0 i} x_{ij}=1$, for a given loop configuration $\mathbf{s}$. More precisely,  
\begin{align}
q_{ij}(x_{ij}=1) &\equiv \left\langle \left( \frac{1 - \xi_{ij} \sigma_i\sigma_j}{2}\right)^{s_{ij}} \right\rangle_0,\\
q_i(\prod_{j\in \bar{\partial}_0 i} x_{ij}=1) &\equiv \left\langle \prod_{j\in \bar{\partial}_0 i} \left(\frac{1-\xi_{ij}\sigma_i\sigma_j}{2} \right)^{s_{ij}} \right\rangle_0.
\end{align}
By $q_i(\prod_{j\in \bar{\partial}_0 i} x_{ij}=1)$ we are indeed taking into account interactions between the neighboring loopy interactions $(ij), (ik)$ with distance $d_{ij,ik}=0$.   

The above approximation for the nonlocal loop interactions can be used in any of the high-temperature loop expansions given in Sec. \ref{S2}.  
In Appendix \ref{app-even} we show how the even or odd parts of expansions \ref{L2} and \ref{L4} can be computed by an approximate message-passing algorithm. In the following, instead we focus on the expression given by Eq. \ref{L5}, where all the expansion terms are nonnegative. 
Then we have
\begin{align}
e^{-\beta(F-F_0-\Delta F_0)} \approx \sum_{\mathbf{s}} \prod_{(ij) \in \mathcal{L}} \tilde{v}_{ij}(s_{ij})\prod_{i \in \mathcal{G}}\tilde{v}_i(s_{\bar{\partial}_0 i}),
\end{align}
where now
\begin{align}
\tilde{v}_{ij}(s_{ij}) \equiv v_{ij}(s_{ij})q_{ij}(x_{ij}=1),\hskip1cm
\tilde{v}_i(s_{\bar{\partial}_0 i}) \equiv \frac{ q_i(\prod_{j\in \bar{\partial}_0 i} x_{ij}=1) }{\prod_{j\in \bar{\partial}_0 i} q_{ij}(x_{ij}=1)}.
\end{align}

The sum over the loop configurations can be computed within the Bethe approximation. Here the cavity marginal $\lambda_{i\to j}(s_{ij})$ in the absence of $(\tilde{v}_{ij},\tilde{v}_{j})$, is recursively obtained by the following BP equations \cite{MM-book-2009}:
\begin{align}\label{bp-cl}
\lambda_{i\to j}(s_{ij}) \propto \sum_{\{s_{ik}|k\in \bar{\partial}_0 i \setminus j\}} \tilde{v}_i(s_{\bar{\partial}_0 i}) \prod_{k\in \bar{\partial}_0 i \setminus j} \left(\tilde{v}_{ik}(s_{ik})\lambda_{k\to i}(s_{ik})\right).
\end{align}
The solutions to these equations are found by iteration; starting from arbitrary cavity messages, we update the messages according to the above equations for a sufficiently large number of iterations to reach a fixed point of the equations. Then, the free energy reads as $F-F_0 -\Delta F_0 \approx \sum_{i\in \mathcal{G}}\Delta F_i-\sum_{(ij)\in \mathcal{L}}\Delta F_{ij}$, where
\begin{align}
e^{-\beta\Delta F_i} &= \sum_{\{s_{ik}|k\in \bar{\partial}_0 i\}} \tilde{v}_i(s_{\bar{\partial}_0 i}) \prod_{k\in \bar{\partial}_0 i} \left(\tilde{v}_{ik}(s_{ik})\lambda_{k\to i}(s_{ik})\right),\\
e^{-\beta\Delta F_{ij}} &= \sum_{s_{ij}}  \tilde{v}_{ij}(s_{ij})\lambda_{j\to i}(s_{ij})\lambda_{i\to j}(s_{ij}).
\end{align}

\begin{figure}
\includegraphics[width=16cm]{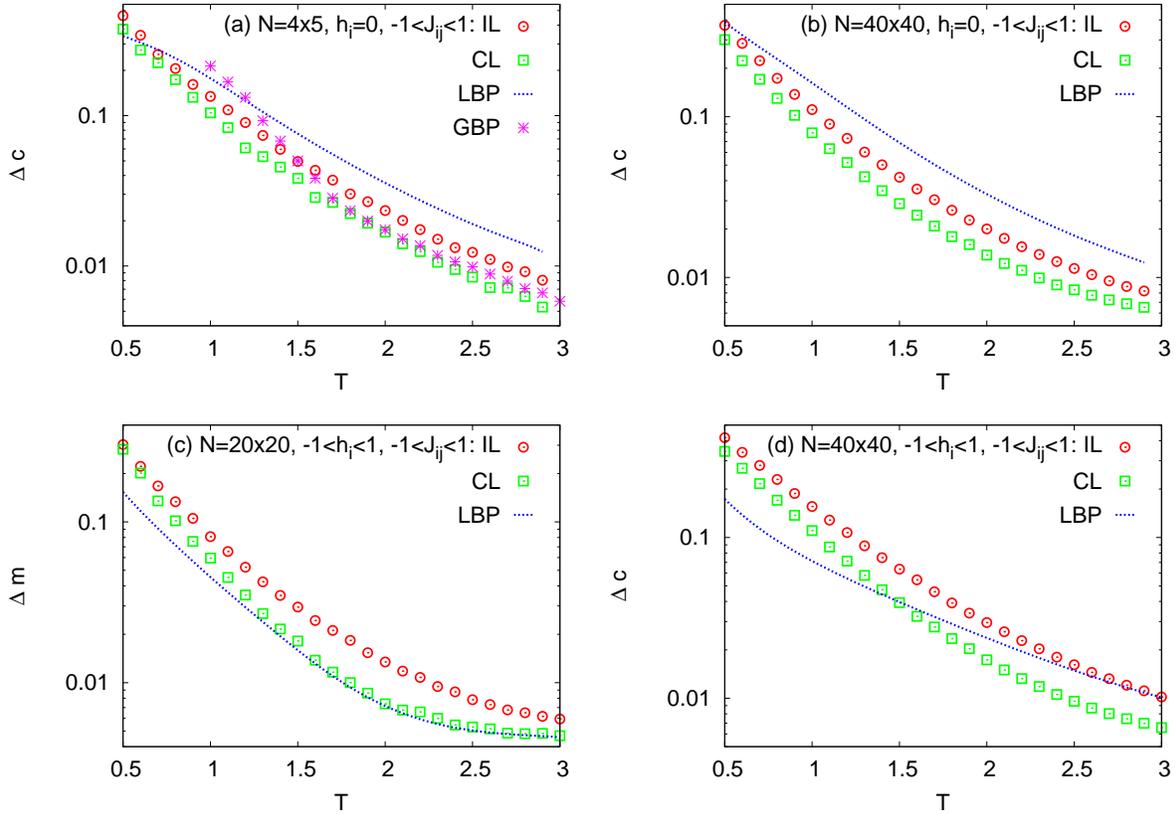}
\caption{(Color online) Comparing the RMS errors in loopy correlations and magnetizations, $\Delta c \equiv \sqrt{\sum_{(ij)\in \mathcal{L}}(c_{ij}^{app}-c_{ij}^{exact})^2/|\mathcal{L}|}$ and $\Delta m \equiv \sqrt{\sum_{i=1}^N(m_i^{app}-m_i^{exact})^2/N}$, obtained by different approximations: independent-loop (IL), correlated-loops (CL), loopy belief propagation (LBP), and generalized belief propagation (GBP) for random Ising models with random couplings uniformly distributed in $(-1,+1)$ on the $2D$ square lattice. The fields are zero in panels (a) and (b), but they are nonzero and uniformly distributed in $(-1,+1)$ in panels (c) and (d). Both the IL and CL approximations are based on a maximum-weight spanning tree, with weight $W=\sum_{(ij)\in \mathsf{T}} |J_{ij}|$.  
The $c_{ij}^{exact}$ and $m_i^{exact}$ are computed by an exact algorithm for the smaller system in panel (a), and by the Monte Carlo algorithm for the larger systems in panels (b)-(d). The errorbars are about the size of the symbols.}\label{f6}
\end{figure}

In Appendix \ref{app-mc} we give the expressions for $m_i$ and $c_{ij}$ obtained by taking derivatives of the above free energy. In particular, we can easily obtain an estimation of spin correlations on loopy interactions, 
\begin{align}
c_{ij} =-\frac{\partial \beta F}{\partial K_{ij}} = -\mathrm{sign}(K_{ij})-\frac{\partial}{\partial K_{ij}}\left( \beta\Delta F_{i}+\beta\Delta F_{j}-\beta\Delta F_{ij} \right), \hskip0.5cm (ij)\in \mathcal{L}. 
\end{align}
Note that in computing the above derivatives, we need only to consider the explicit dependence of the free energy on $K_{ij}$, because the Bethe free energy is stationary with respect to changes in the cavity messages at the fixed point.
Figure \ref{f6} displays the error in spin correlations on loopy interactions obtained by the above correlated-loops (CL) approximation for some random Ising models on the two-dimensional square lattice. For comparison, the figure also shows the error obtained by the IL approximation, loopy BP algorithm, and generalized BP (shown only for the small system in panel (a)). The latter works with larger regions of variables (here plaquettes) to extend the consistency range of the local marginals \cite{YFW-nips-2001,P-jphysa-2005,GBP-jstat-2011}. We see that already the IL approximation is working better than the loopy BP in the absence of the external fields. And, considering the interactions between the neighboring loop variables in the CL approximation results in spin correlations that are comparable to those of the generalized BP. Nevertheless, one could still obtain better results with the loopy BP for small temperatures, especially in the presence of strong fields. The same is true for homogeneous systems with uniform couplings, where the absence of a dominant spanning tree means we have to take into account the longer-ranged interactions between the loop variables.

We can still improve on the above approximation by increasing the interaction range of the loop variables.
For instance, to get better approximations, we may add interactions between the more distant loop variables, 
\begin{multline}
P_0\left(\prod_{(ij)\in \mathcal{L}} x_{ij}=1\right) \approx \prod_{(ij) \in \mathcal{L}} q_{ij}(x_{ij}=1) \prod_{i \in \mathcal{G}}\frac{ q_i(\prod_{j\in \bar{\partial}_0 i} x_{ij}=1) }{\prod_{j\in \bar{\partial}_0 i} q_{ij}(x_{ij}=1)}\\
\times \prod_{(ij)<(kl) \in \mathcal{L}} \frac{q_{ij,kl}(x_{ij}x_{kl}=1) }{q_{ij}(x_{ij}=1)q_{kl}(x_{kl}=1)},
\end{multline}
where 
\begin{align}
q_{ij,kl}(x_{ij}x_{kl}=1) \equiv \left\langle (\frac{1-\xi_{ij}\sigma_i\sigma_j}{2})^{s_{ij}}(\frac{1-\xi_{kl}\sigma_k\sigma_l}{2})^{s_{kl}}  \right\rangle_0. 
\end{align}
The interactions $q_{ij,kl}(x_{ij}x_{kl}=1)$ can be limited to close loopy interactions with distances $d_{ij,kl}$ smaller than a given distance $d^*$. Again, from the clustering property, we expect to obtain a good approximation already for small $d^*$. We leave further studies in this direction for future works.

\section{Conclusion}\label{S4}
We started from a high-temperature loop expansion for the Ising model on a loopy interaction graph regarded as a globally interacting system of loop variables. We then suggested different approximations for the nonlocal loop interactions, writing the problem of computing loop corrections in a form that is amenable to local message-passing algorithms. The quality of these approximations depends on the structure of the loopy interactions; when the loopy interactions have a tree structure, we can use the belief propagation algorithm to compute "exactly" the loop corrections within a local approximation of the nonlocal loop interactions. This provides approximate message-passing algorithms that rely on the structure of the loopy interactions, and thus exhibiting better convergence properties than the loopy BP algorithm. In this way, we could obtain more accurate estimations for spin correlations on loopy interactions from these approximations than from the loopy BP algorithm for the Ising model on the $2D$ square lattice with random couplings.     

It would be interesting to find out the typical loop configurations close to a phase transition point. The relevant loop configurations with no frustration  could behave differently as we approach ferromagnetic or spin-glass phase transition points. Close to a continuous phase transition the loop variables are strongly correlated, and the global nature of loop interactions becomes more important. Therefore, it would be important to have better approximations for the nonlocal loop interactions, and work with the higher-order interactions between the loop variables. Finally, instead of starting from a spanning tree in the loop expansions, we could start from any subgraph that allows for accurate computation of the local expectations needed in the expansions. For instance, the  starting subgraph can be obtained by adding an appropriate subset of loopy interactions to the spanning tree \cite{R-pre-2013}.

\acknowledgments We would like to thank A. Braunstein, E. Daryaei, and J. Realpe-Gomez for reading the manuscript and helpful discussions.

\appendix

\section{Computing the local magnetizations and correlations}\label{app-mc}
Given the free energy $F$, we obtain the local expectation values from $m_i=\left\langle \sigma_i \right\rangle=-\frac{\partial \beta F}{\partial B_{i}}$ and $c_{ij}=\left\langle \sigma_i\sigma_j \right\rangle=-\frac{\partial \beta F}{\partial K_{ij}}$.

In the independent-site (IS) approximation the free energy is given by $F\approx F_0 + \sum_{i\in \mathcal{G}}\Delta F_i-\sum_{(ij)\in \mathcal{L}}\Delta F_{ij}$, where $F_0$ is the free energy of the system living on the spanning tree. The other terms give the loop corrections computed by the belief propagation (BP) algorithm in the IS approximation. The precise definitions along with the computation of the local magnetizations can be found in the main text, Sec. \ref{S32}. For the two-spin correlations, we have
\begin{align}
c_{ll'} &\approx  -\frac{\partial }{\partial K_{ll'}}\left(\beta\Delta F_l+ \beta\Delta F_{l'}-\beta\Delta F_{ll'}\right), \hskip0.5cm (ll')\in \mathcal{L}\\
c_{ll'} &\approx \left\langle \sigma_l\sigma_{l'} \right\rangle_0 - \sum_{i\in \mathcal{G}}\frac{\partial \beta\Delta F_i}{\partial K_{ll'}}, \hskip0.5cm (ll')\notin \mathcal{L}. 
\end{align}
To compute the above quantities we need to know the BP messages $\lambda_{i\to j}(s_{ij})$ satisfying Eq. \ref{bp-is}, and the following derivatives:
\begin{align}
\frac{\partial u_{ll'}(s_{ll'})}{\partial K_{ll'}} &=\delta_{s_{ll'},0}\sinh K_{ll'}+\delta_{s_{ll'},1}\cosh K_{ll'}, \hskip0.5cm (ll')\in \mathcal{L}\\
\frac{\partial u_i(s_{\bar{\partial}_0 i})}{\partial K_{ll'}} &=\left( \frac{1-(-1)^{\sum_{j\in \bar{\partial}_0 i}s_{ij}}}{2}\right)\left( \left\langle \sigma_i\sigma_l\sigma_{l'}\right\rangle_0-\left\langle \sigma_i\right\rangle_0\left\langle \sigma_l\sigma_{l'}\right\rangle_0 \right), \hskip0.5cm (ll')\notin \mathcal{L}. 
\end{align}
Here the loop variables $s_{ij}=0,1$ show the absence or presence of loopy interactions, and the averages $\langle \cdots \rangle_0$ are taken in the tree interaction graph. 

In the correlated-loops (CL) approximation, the free energy reads $F\approx F_0+\Delta F_0 + \sum_{i\in \mathcal{G}}\Delta F_i-\sum_{(ij)\in \mathcal{L}}\Delta F_{ij}$. The precise definitions are given in the main text, Sec. \ref{S33}. Here we obtain a local magnetization by
\begin{align}
m_l=-\frac{\partial \beta F}{\partial B_l} \approx \left\langle \sigma_l \right\rangle_0 + \sum_{i\in \mathcal{G}}\delta m_i^l-\sum_{(ij)\in \mathcal{L}}\delta m_{ij}^l, 
\end{align}
where
\begin{align}
\delta m_i^l &\equiv \frac{\sum_{\{s_{ik}|k\in \bar{\partial}_0 i\}} q_i^l(\prod_{j\in \bar{\partial}_0 i} x_{ij}=1) \prod_{k\in \bar{\partial}_0 i} \left(v_{ij}(s_{ij})\lambda_{k\to i}(s_{ik})\right)}{\sum_{\{s_{ik}|k\in \bar{\partial}_0 i\}} \tilde{v}_i(s_{\bar{\partial}_0 i}) \prod_{k\in \bar{\partial}_0 i} \left(\tilde{v}_{ik}(s_{ik})\lambda_{k\to i}(s_{ik})\right)},\\
\delta m_{ij}^l &\equiv \frac{\sum_{s_{ij}}  v_{ij}(s_{ij})q_{ij}^l(x_{ij}=1)\lambda_{j\to i}(s_{ij})\lambda_{i\to j}(s_{ij})}{\sum_{s_{ij}}  \tilde{v}_{ij}(s_{ij})\lambda_{j\to i}(s_{ij})\lambda_{i\to j}(s_{ij})}. 
\end{align}
To compute the loop corrections, we need the cavity messages $\lambda_{i\to j}(s_{ij})$ governed by Eq. \ref{bp-cl}, and the following derivatives:
\begin{align}
q_i^l &\equiv \frac{\partial q_i}{\partial B_l}=\left\langle \sigma_l \prod_{j\in \bar{\partial}_0 i} (\frac{1-\xi_{ij}\sigma_i\sigma_j}{2})^{s_{ij}}\right\rangle_0- \left\langle \sigma_l\right\rangle_0\left\langle \prod_{j\in \bar{\partial}_0 i} (\frac{1-\xi_{ij}\sigma_i\sigma_j}{2})^{s_{ij}}\right\rangle_0,\\
q_{ij}^l &\equiv \frac{\partial q_{ij}}{\partial B_l}=\left\langle \sigma_l(\frac{1-\xi_{ij}\sigma_i\sigma_j}{2})^{s_{ij}}\right\rangle_0-\left\langle \sigma_l\right\rangle_0\left\langle (\frac{1-\xi_{ij}\sigma_i\sigma_j}{2})^{s_{ij}}\right\rangle_0.
\end{align}

The two-spin correlations in the CL approximation are given by
\begin{align}
c_{ll'} &\approx -\mathrm{sign}(K_{ll'})-\frac{\partial}{\partial K_{ll'}}\left( \beta\Delta F_{l}+\beta\Delta F_{l'}-\beta\Delta F_{ll'} \right), \hskip0.5cm (ll')\in \mathcal{L}\\ \label{cij-CL}
c_{ll'} &\approx \left\langle \sigma_l\sigma_{l'} \right\rangle_0 + \sum_{i\in \mathcal{G}}\delta c_i^{ll'}-\sum_{(ij)\in \mathcal{L}}\delta c_{ij}^{ll'}, \hskip0.5cm (ll')\notin \mathcal{L}, 
\end{align}
where 
\begin{align}
\delta c_i^{ll'} &\equiv \frac{\sum_{\{s_{ik}|k\in \bar{\partial}_0 i\}} q_i^{ll'}(\prod_{j\in \bar{\partial}_0 i} x_{ij}=1) \prod_{k\in \bar{\partial}_0 i} \left(v_{ij}(s_{ij})\lambda_{k\to i}(s_{ik})\right)}{\sum_{\{s_{ik}|k\in \bar{\partial}_0 i\}} \tilde{v}_i(s_{\bar{\partial}_0 i}) \prod_{k\in \bar{\partial}_0 i} \left(\tilde{v}_{ik}(s_{ik})\lambda_{k\to i}(s_{ik})\right)},\\
\delta c_{ij}^{ll'} &\equiv \frac{\sum_{s_{ij}}  v_{ij}(s_{ij})q_{ij}^{ll'}(x_{ij}=1)\lambda_{j\to i}(s_{ij})\lambda_{i\to j}(s_{ij})}{\sum_{s_{ij}}  \tilde{v}_{ij}(s_{ij})\lambda_{j\to i}(s_{ij})\lambda_{i\to j}(s_{ij})}. 
\end{align}
In the above equations, we defined 
\begin{align}
q_i^{ll'} &\equiv \frac{\partial q_i}{\partial K_{ll'}}=\left\langle \sigma_l\sigma_{l'} \prod_{j\in \bar{\partial}_0 i} (\frac{1-\xi_{ij}\sigma_i\sigma_j}{2})^{s_{ij}}\right\rangle_0- \left\langle \sigma_l\sigma_{l'}\right\rangle_0\left\langle \prod_{j\in \bar{\partial}_0 i} (\frac{1-\xi_{ij}\sigma_i\sigma_j}{2})^{s_{ij}}\right\rangle_0,\\
q_{ij}^{ll'} &\equiv \frac{\partial q_{ij}}{\partial K_{ll'}}=\left\langle \sigma_l\sigma_{l'}(\frac{1-\xi_{ij}\sigma_i\sigma_j}{2})^{s_{ij}}\right\rangle_0-\left\langle \sigma_l\sigma_{l'}\right\rangle_0\left\langle (\frac{1-\xi_{ij}\sigma_i\sigma_j}{2})^{s_{ij}}\right\rangle_0.
\end{align}

\section{Computing the even free-energy contribution $\Delta F_{even}^+$}\label{app-even}
In this section we explain an approximate message-passing algorithm to compute the contribution of loop configurations $\mathbf{s}=\{s_{ij}=0,1|(ij) \in \mathcal{L}\}$ with an even number of ferromagnetic loopy interactions,
\begin{align}
e^{-\beta \Delta F_{even}^+} &= \sum_{\mathbf{s}:M_+=even} \prod_{(ij) \in \mathcal{L}} |v_{ij}^+(s_{ij})| \left\langle \prod_{(ij) \in \mathcal{L}}(\frac{1-\sigma_i\sigma_j}{2})^{s_{ij}}  \right\rangle_0.
\end{align}
Here $M_+(\mathbf{s})\equiv \sum_{(ij) \in \mathcal{L}}s_{ij}\theta(K_{ij})$, $\theta(x)$ is the Heaviside function, and $v_{ij}^+(s_{ij})=(e^{-2K_{ij}}-1)$.
The aim is to find an estimation of $\Delta F_{even}^+$ by the Bethe approximation, given an approximate factorization of the nonlocal loop interactions 
\begin{align}
\left\langle \prod_{(ij) \in \mathcal{L}}(\frac{1-\sigma_i\sigma_j}{2})^{s_{ij}}  \right\rangle_0 \approx \prod_{(ij) \in \mathcal{L}} q_{ij}(x_{ij}=1) \prod_{i \in \mathcal{G}}\frac{ q_i(\prod_{j\in \bar{\partial}_0 i} x_{ij}=1) }{\prod_{j\in \bar{\partial}_0 i} q_{ij}(x_{ij}=1)}.
\end{align}
Here $x_{ij}= (\frac{1-\sigma_i\sigma_j}{2})^{s_{ij}}$, and $q_{ij}(x_{ij}=1)$ is the probability of having $x_{ij}=1$ in the tree interaction graph.  Similarly, $q_i(\prod_{j\in \bar{\partial}_0 i} x_{ij}=1)$ is the probability of having $\prod_{j\in \bar{\partial}_0 i} x_{ij}=1$. Then we have
\begin{align}
e^{-\beta \Delta F_{even}^+} \approx \sum_{\mathbf{s}:M_+=even} \prod_{(ij) \in \mathcal{L}} \tilde{v}_{ij}(s_{ij})\prod_{i \in \mathcal{G}}\tilde{v}_i(s_{\bar{\partial}_0 i}),
\end{align}
with
\begin{align}
\tilde{v}_{ij}(s_{ij}) =|v_{ij}^+(s_{ij})|q_{ij}(x_{ij}=1),\hskip1cm
\tilde{v}_i(s_{\bar{\partial}_0 i}) =\frac{ q_i(\prod_{j\in \bar{\partial}_0 i} x_{ij}=1) }{\prod_{j\in \bar{\partial}_0 i} q_{ij}(x_{ij}=1)}.
\end{align}

Let us for simplicity assume that the graph $\mathcal{G}$ induced by the loopy interactions is a tree, otherwise we choose one of its spanning trees \cite{RRZ-epjb-2011,RZ-prb-2012}.
Then, we introduce auxiliary variables $\tau_{i\to j}\in \{-1,+1\}$ on the directed links of $\mathcal{G}$, where $\tau_{i\to j}=(-1)^{M_+^{i\to j}(\mathbf{s})}$ gives the parity of the ferromagnetic loopy interactions in the cavity tree $\mathcal{G}_{i\to j}$; a branch of the tree that includes $i$ and cavity trees $\{\mathcal{G}_{k\to i}|k \in \bar{\partial}_0 i \setminus j\}$. More precisely, 
\begin{align}
\tau_{i\to j}=(-1)^{\sum_{k\in \bar{\partial}_0 i \setminus j} s_{ki}\theta(K_{ki})}\prod_{k\in \bar{\partial}_0 i \setminus j}\tau_{k\to i} \equiv \hat{\tau}_{i\to j}.
\end{align}
Now we have to work with extended marginals $\lambda_{i\to j}(s_{ij},\tau_{i\to j})$ governed by the following BP equations
\begin{align}
\lambda_{i\to j}(s_{ij},\tau_{i\to j}) \propto \sum_{\{s_{ik},\tau_{k\to i}|k\in \bar{\partial}_0 i \setminus j\}}\delta(\tau_{i\to j}-\hat{\tau}_{i\to j}) \tilde{v}_i(s_{\bar{\partial}_0 i}) \prod_{k\in \bar{\partial}_0 i \setminus j} \left(\tilde{v}_{ik}(s_{ik})\lambda_{k\to i}(s_{ik},\tau_{k\to i})\right).
\end{align}
The solution to these equations can be found by iteration. Then, for an even number of ferromagnetic loopy interactions, the free energy shifts are given by
\begin{align}
e^{-\beta\Delta F_i} &= \sum_{\{s_{ik}, \tau_{k\to i}|k\in \bar{\partial}_0 i\}} \delta(\tau_{i}-1)\tilde{v}_i(s_{\bar{\partial}_0 i}) \prod_{k\in \bar{\partial}_0 i} \left(\tilde{v}_{ik}(s_{ik})\lambda_{k\to i}(s_{ik})\right),\\
e^{-\beta\Delta F_{ij}} &= \sum_{s_{ij},\tau_{i\to j},\tau_{j\to i}} \delta(\tau_{ij}-1) \tilde{v}_{ij}(s_{ij})\lambda_{j\to i}(s_{ij},\tau_{j\to i})\lambda_{i\to j}(s_{ij},\tau_{i\to j}),
\end{align}
where $\tau_{ij}\equiv (-1)^{s_{ij}\theta(K_{ij})}\tau_{i\to j}\tau_{j\to i}$ and the local $\tau_i$ is computed like the cavity one $\tau_{i\to j}$ but considering the contribution from all the neighbors in $\bar{\partial}_0 i$. Finally, the even free energy reads as $\Delta F_{even}^+=\sum_{i}\Delta F_i-\sum_{(ij)\in \mathcal{L}}\Delta F_{ij}$.

\end{document}